\documentclass[a4paper]{jpconf}
\usepackage{xcolor}

\usepackage{fullpage}
\usepackage{graphicx}
\usepackage{pgfplots}
\usepackage{subcaption}
\usepackage{amssymb,amsmath}
\usepackage[unicode=true]{hyperref}
\usepackage[utf8x]{inputenc}
\usepackage[T1]{fontenc}
\usepackage{bbm}
\usepackage{hyperref}

\begin{document}

\title{Conserved quantities, exceptional points, and antilinear symmetries in non-Hermitian systems}
\author{Frantisek Ruzicka$^{1\dagger}$, Kaustubh S. Agarwal$^2$ and Yogesh N. Joglekar$^{2\ddagger}$}
\address{$^1$Institute  of  Nuclear  Physics,  Czech  Academy  of  Sciences,  Rez  250  68,  Czech  Republic\\ 
$^2$Department  of  Physics,  Indiana  University  Purdue  University Indianapolis  (IUPUI),  Indianapolis,  Indiana  46202, United States}
\ead{$^\dagger$fruzicka@gmail.com, $^\ddagger$yojoglek@iupui.edu}

\begin{abstract}
Over the past two decades, open systems that are described by a non-Hermitian Hamiltonian have become a subject of intense research. These systems encompass classical  wave systems with balanced gain and loss, semiclassical models with mode selective losses, and minimal quantum systems, and the meteoric research on them has mainly focused on the wide range of novel functionalities they demonstrate. Here, we address the following questions: Does anything remain constant in the dynamics of such open systems? What are the consequences of such conserved quantities? Through spectral-decomposition method and explicit, recursive procedure, we obtain all conserved observables for general $\mathcal{PT}$-symmetric systems. We then generalize the analysis to Hamiltonians with other antilinear symmetries, and discuss the consequences of conservation laws for open systems. We illustrate our findings with several physically motivated examples. 
\end{abstract}

\section{Introduction}
Discrete or continuous symmetries and conservation laws that result from them have been instrumental to the theoretical and experimental developments in physics over the past few centuries~\cite{Goldstein2002,Landau1976}. Identifying the quantities that remain constant during temporal evolution of a system allows us to severely constraint their  global dynamics. Starting from Kepler's law (equal-area-swept in equal-times for planar orbits)~\cite{Feynman2013}, these time-invariants have underpinned numerous scientific advances; they include the prediction of (electron) neutrinos based on the requirement of energy, momentum, and angular momentum conservation in beta-decay processes~\cite{Close2010}. On the theoretical front, conservation laws have played critical role in the development of approximate methods that are indispensable for studying interacting, many-body systems. These self-consistent or conserving approximations characterize a special set of Feynman diagrams, used to calculate response functions or susceptibilities, such that the approximate answers still obey relevant conservation laws~\cite{Baym1961,Mahan2000}. 

The dynamics of isolated systems are governed by static Hamiltonians. For classical systems, this Hamiltonian (energy) function, defined over the phase space, is real. In this case, all conserved quantities are those whose Poisson bracket with the Hamiltonian vanishes, i.e. they commute with the Hamiltonian~\cite{Goldstein2002}. For quantum systems, the same criterion is applicable although now the Hamiltonian $H_0$ is a Dirac-Hermitian operator, $H_0=H_0^\dagger$, and so are ``observables" in the quantum theory~\cite{Sakurai2011}. A full set of conserved quantities for such a system are therefore obtained by identifying all independent observables that obey the commutation property.  Some of the conserved observables are immediately apparent. An obvious one is the identity operator, $\mathbbm{1}$, that gives rise to the conservation of the Dirac inner-product between two states, and in particular, the norm of a given state. A second observable is the Hamiltonian itself, and it expresses the conservation of energy for a given, isolated system. In both classical and quantum cases, the dynamics generated by a Hamiltonian are reversible and thus do not evolve into a steady state. 

In contrast, the dynamics of a classical system coupled to an environment become irreversible in the thermodynamic limit, and the system evolves towards the ground state or a non-equilibrium steady state. In such cases, although conservation laws apply to the system+environment, the existence or implications of time-invariant quantities for a classical system alone are unclear. When a small quantum system is coupled to an environment, it loses its coherence. This process is traditionally described by the Gorini-Kossakowski-Sudarshan Lindblad equation~\cite{Sudarshan1961,Gorini1976,Lindblad1976} for the (reduced) density matrix of the system, and results in a completely-positive trace preserving (CPTP) map on the space of physical density matrices~\cite{Manzano2020}. In such cases, the conserved quantities are determined by observables that simultaneously commute with the system Hamiltonian and all relevant Lindblad dissipators~\cite{Albert2014}. Absent special symmetries, it implies that the {\it state-norm or the trace of the system density-matrix is the only conserved quantity} for such quantum channels. 

In this article, we will review symmetries and conservation laws for a new class of systems that are governed by non-Hermitian Hamiltonians~\cite{Bian2020}. Their coherent, non-unitary, non-norm-preserving dynamics are qualitatively different from the coherent, unitary dynamics of isolated systems as well as the incoherent, non-unitary, trace-preserving dynamics of systems coupled to an environment. Since Bender and coworkers' discovery~\cite{Bender1998,Bender2001}, over the past two decades, non-Hermitian Hamiltonians that are invariant under combined operations of parity and time-reversal ($\mathcal{PT}$) have emerged as new research frontier~\cite{Bender2007}. Its meteoric growth was driven first by theoretical interest in the generalization of a quantum theory to non-Hermitian, self-adjoint Hamiltonians~\cite{Mostafazadeh2002,Mostafazadeh2010}, and then by experiments on open, classical wave systems~\cite{Joglekar2013} with balanced gain and loss that are faithfully described by $\mathcal{PT}$-symmetric Hamiltonians~\cite{ElGanainy2018}. A prototypical $\mathcal{PT}$-symmetric Hamiltonian $H(\gamma)\neq H^\dagger(\gamma)$ has a purely real spectrum when the gain-loss strength $\gamma$ is smaller than a threshold value $\gamma_\mathrm{PT}$. At $\gamma=\gamma_\mathrm{PT}$, two (or more) eigenvalues of $H(\gamma)$ coincide and the corresponding eigenvectors coalesce, giving rise to an exceptional point (EP) degeneracy~\cite{Kato1995,Ozdemir2019}. In contrast to the diabolic point (DP) degeneracies of self-adjoint operators, the eigenvectors of $H(\gamma_\mathrm{PT})$ do not span the space. When $\gamma$ exceeds the threshold, the degenerate eigenvalues of $H(\gamma)$ turn into complex conjugate pairs and the non-orthogonal eigenvectors of $H(\gamma)$ span the space again. This transition from real to complex-conjugate eigenvalues is called $\mathcal{PT}$-symmetry breaking transition. We remind the reader that the time-evolution operator $G(t)\equiv\exp[-iH(\gamma)t)]$ ($\hbar=1$) is not unitary due to the non-orthogonal nature of the eigenstates of $H(\gamma)$ irrespective of whether the spectrum of $H(\gamma)$ is real or complex-conjugate pairs. In physical terms, the system changes from having purely oscillatory eigenmodes and a bounded, oscillatory state-norm to amplifying and decaying eigenmodes that, in the absence of nonlinearities, lead to an exponentially growing state-norm. Given this qualitatively different behaviors across the $\mathcal{PT}$-symmetry breaking transition at the EP, it makes sense to ask, ``Is there anything that remains time-invariant in these open systems with balanced gain and loss''? As we will show below, this question can be completely answered with a recursive, analytical construction. 

After the extensive realizations of $\mathcal{PT}$-symmetric Hamiltonians in classical platforms that include coupled mechanical~\cite{Bender2013} and electrical~\cite{Schindler2011,Schindler2012,Chitsazi2017,LeonMontiel2018,Wang2020} oscillators, coupled optical waveguides~\cite{Duchesne2009,Ruter2010}, fiber loops~\cite{Regensburger2012,Weidemann2020}, microring resonators~\cite{Peng2014}, acoustics~\cite{Zhu2014}, optomechanical systems~\cite{Lu2015}, and so on, past two years have seen tremendous efforts to extend these ideas into the semiclassical or purely quantum domains. These efforts have led to the realizations of passive $\mathcal{PT}$-symmetric Hamiltonians with mode-selective losses in ultracold atoms in a two-level system~\cite{Li2019} and momentum-space lattice~\cite{Chen2020}, a single NV center~\cite{Wu2019}, single~\cite{Xiao2019} and correlated photons~\cite{Klauck2019}, and a superconducting transmon circuit~\cite{Naghiloo2019}. As we will show below, although these quantum systems are governed by lossy Hamiltonians, conservation laws apply to them as well and lead to observable consequences that arise from the difference between decay rate for the slow mode that emerges after the passive $\mathcal{PT}$-symmetry breaking transition~\cite{Joglekar2018} and the average decay rate for the lossy system. 

The plan for the paper is as follows. In Sec.~\ref{sec:theory} we present an algebraic approach to obtain conserved observables and generalize the results from those for a $\mathcal{PT}$-symmetric Hamiltonian to Hamiltonians with a broader class of symmetries. We show that the nature of the conservation laws is purely algebraic, not restricted to the Hermiticity of the system in any way. Instead, it depends heavily on the locations of the eigenvalues, itself is a reflection of some underlying symmetry of the system. We prove existence conditions for the conserved quantities. In Sec.~\ref{sec:recursive}, we review a recursive procedure that allows explicit calculation of the conserved, intertwining operators without using the spectral decomposition approach of Sec.~\ref{sec:theory}, and illustrate our construction with a few examples. We generalize the ideas of conserved quantities to non-$\mathcal{PT}$-symmetric Hamiltonians in Sec.~\ref{sec:antipt}. Section~\ref{sec:conc} concludes the paper with remarks about consequences for lossy $\mathcal{PT}$-symmetric systems, generalization to time-periodic, non-Hermitian systems and systems undergoing non-unitary, discrete-time quantum walks, and a summary.

As an aside, we note that in this work, we {\it only consider Dirac-Hermiticity as the defining criterion} for all observables; this is due to fact that nature seems only consistent with a Born-rule in the quantum theory that uses Dirac Hermitian-conjugate to obtain the probability density. Therefore, we will not consider formally consistent mathematical models with non-Hermitian Hamiltonians or ``observables'' that are self-adjoint~\cite{Mostafazadeh2010,Znojil2009,Znojil2015}. We will also restrict ourselves to finite-dimensional systems, thereby circumventing the issues of domains, boundedness, and invertibility~\cite{Mostafazadeh2020}. 


\section{Conserved quantities in non-Hermitian systems}
\label{sec:theory}

In classical mechanics, observables that are conserved during time evolution are most easily defined in the Hamiltonian formalism, where they are determined by a vanishing Poisson bracket~\cite{Goldstein2002}. Therefore, it should come at no surprise that their quantum counterparts become by far most apparent in the Heisenberg picture. Let the system under consideration be governed by a static Hamiltonian $H$ that may not be Hermitian.  It satisfies equations of motion $i\partial_t |\psi(t)\rangle=H|\psi(t)\rangle$ and $-i\partial_t\langle \psi(t)|=\langle\psi(t)|H^\dagger$. By definition, a linear operator $\eta$ is a constant of motion if and only if $\langle\psi(t)|\eta|\psi(t)\rangle=\mathrm{Tr}[\eta\rho_\psi(t)]$ remains constant for any arbitrary state $|\psi\rangle$ (or a density matrix $\rho_\psi$). In the absence of intrinsic time-dependence, this constraint translates into
\begin{equation}
\label{eq:eta1}
i\frac{d}{dt}\langle\psi(t)|\eta|\psi(t)\rangle=\langle\psi(t)|\eta H-H^\dagger\eta|\psi(t)\rangle=0.
\end{equation}
Due to the linearity of the constraint in Eq.(\ref{eq:eta1}), without loss of generality, we can choose $\eta$ to be a Hermitian matrix. When $H=H^\dagger$ (an isolated system), the observable conservation is therefore equivalent to commutation, just as expected. A Hermitian system trivially leaves two important operators conserved, namely the identity (state-norm) and the Hamiltonian itself (energy). We emphasize that many more independent, conserved operators can be constructed very easily; an example is the set of Hermitian projectors onto the eigenspace for each real eigenvalue. These are often disregarded in many applications since they are not connected to fundamental symmetries of the system. However, they fulfill all requirements for a genuine conserved observable and will play an important role in the following analysis. When $H$ is not Hermitian, Eq.(\ref{eq:eta1}) leads to the following intertwining constraint, \begin{equation}
\label{eq:eta2}
\eta H=H^\dagger \eta.
\end{equation}

This characterization of conserved observables as intertwining operators, Eq.(\ref{eq:eta2}), has appeared in the literature in the context of pseudo-Hermitian operators ($\eta$ is invertible)~\cite{Mostafazadeh2002,Mostafazadeh2010,Znojil2009} and the $\mathcal{CPT}$-inner product ($\eta$ is positive definite)~\cite{Bender2001}. Here, we only focus on them as conserved observables for an open system with gain and loss. 


\subsection{Conserved quantities via spectral decomposition}
\label{subsec:spectral}

To show the existence of conserved operators, we proceed by direct construction using spectral decomposition techniques~\cite{Mostafazadeh2002}. It is a simple observation that we can restrict our attention to $n$-dimensional Hermitian operators in the following analysis. Since they form the basis of the underlying operator Hilbert space of dimension $n^2$, they can serve as generators of all the other, possibly non-Hermitian, operators by complex linear combinations. 

For a given $n$-dimensional Hamiltonian $H$ and an unknown $\eta$, solving the intertwining relation Eq.(\ref{eq:eta2}) is equivalent to solving a set of $n^2$ linear equations with $n^2$ unknowns. The question posed here---identification of all conserved quantities---concerns the dimension of its solution space. For this section, we will assume that the Hamiltonian $H$ with non-degenerate eigenvalues $\epsilon_k$ has a complete set of right eigenvectors $\{|R_k \rangle\}$ defined by $H|R_k\rangle=\epsilon_k|R_k\rangle$ and let $\{\langle L_k|\}$ be the left-eigenvectors that satisfy $\langle L_k|H=\epsilon_k\langle L_k|$. In the spirit of choosing our conventions close to the experiments, we will normalize these eigenvectors according to the Dirac norm, $\langle R_k|R_k\rangle=1=\langle L_m|L_m\rangle$. We also remind the reader that right-eigenvectors are not orthogonal to each other, $\langle R_k|R_m\rangle\neq 0$, and neither are the left eigenvectors. However, they form a biorthogonal basis, $\langle L_k|R_m\rangle\propto\delta_{km}$. We note here that the general case where the eigenvalue $\epsilon_m$ has a degeneracy $d_m$ has been discussed, albeit with different normalization, in Ref.~ \cite{Mostafazadeh2002}. The Hamiltonian $H\neq H^\dagger$ admits a spectral decomposition 
\begin{eqnarray}
\label{eq:h1}
H = \sum_{k=1}^N \epsilon_k \frac{| R_k \rangle \langle L_k |}{\langle L_k|R_k\rangle} &,&
H^\dagger = \sum_{k=1}^N \epsilon^*_k \frac{| L_k \rangle \langle R_k |}{\langle R_k|L_k\rangle} 
\end{eqnarray}
in such a basis where the denominator $\langle L_k|R_k\rangle\neq 1$ due to our choice of Dirac normalization.  Equation~(\ref{eq:eta1}) implies that the eigenvalues of $H$ are either purely real or occur in complex conjugate pairs for a nonzero intertwining operator to exist. This means the characteristic equation for eigenvalues of $H$ has real coefficients or, equivalently, there is an antilinear operator $\mathcal{A}$ that commutes with $H$~\cite{Mostafazadeh2002}. Without loss of generality, we will call  this antilinear operator $\mathcal{A}=\mathcal{PT}$. From the definitions, it also follows that $\eta|R_k\rangle\propto|L_k\rangle$ provided $\epsilon_k$ is real, and $\eta|R_\alpha\rangle\propto |L_{\alpha^*}\rangle$ provided $\epsilon_{\alpha^*}\neq\epsilon_\alpha$ is a complex energy. By sandwiching the intertwining relation between $\langle R_k|$ and $|R_j\rangle$, we also get $(\epsilon_j-\epsilon^*_k)\langle R_k|\eta|R_j\rangle=0$~\cite{Mostafazadeh2002}. It is now easy to explicitly show that the following Hermitian operators
\begin{align}
\label{eq:eta10}
\eta_k& =|L_k\rangle\langle L_k|,\\
\label{eq:eta11}
\eta_{s\alpha}& = (|L_\alpha\rangle\langle L_{\alpha^*}|+|L_{\alpha^*}\rangle\langle L_\alpha|)/2,\\
\label{eq:eta12}
\eta_{a\alpha}& = i( |L_\alpha\rangle\langle L_{\alpha^*}|-|L_{\alpha^*}\rangle\langle L_\alpha|)/2,
\end{align}
are intertwining operators. Here the index $k$ ($\alpha,\alpha^*$) spans over real (complex conjugate) eigenvalues. These are analogues of conserved eigenvector projections in traditional Hermitian quantum mechanics, but their physical meaning is certainly much more subtle. Further analysis of this equation will depend on the degeneracy of the spectrum of the Hamiltonian under investigation. We are going to separately address the cases of non-degenerate spectrum, pure eigenvalue degeneracy and the exceptional points.


\subsection{Non-degenerate spectrum}
\label{subsec:nond}

If the Hamiltonian under consideration has a non-degenerate spectrum, there are no more remaining solutions to be found.  We can thus conclude that the number of conserved operators is equal to the number of eigenvalues of $H$, i.e. the count of independent conserved quantities reaches $n$. The completely general formula for $\eta$ in such a case is given by 
\begin{equation}
\eta = \sum_{k\in\mathrm{real}}A_k\eta_k+\sum_{\alpha\in\mathrm{complex}}(A_{s\alpha}\eta_{s\alpha}+A_{a\alpha}\eta_{a\alpha}). 
\end{equation}

As a concrete example, let us consider $H_3=JS_x+i\gamma S_z$ where $S_x$ and $S_z$ are the 3-dimensional representation of SU(2)~\cite{Graefe2008}. This non-Hermitian Hamiltonian commutes with $\mathcal{P_3T}$ operator where $\mathcal{P}_3=\mathrm{antidiag}(1,1,1)$ and $\mathcal{T}=*$, the complex-conjugation operator. Classically, the Hamiltonian $H_3$ represents a $\mathcal{PT}$-symmetric trimer and it has a third-order EP at $\gamma_\mathrm{EP}=J$~\cite{Hodaei2017}. Since $H_3$ is transpose-symmetric, its left-eigenvectors are just transpose of the right ones, i.e. $\langle L_k|=|R_k\rangle^T$. When $\gamma/J=\sin\theta\leq 1$, the purely real eigenvalues of $H_3$ are given by $\epsilon_k=\{\pm\cos\theta,0\}$ and the corresponding Dirac-normalized, right eigenvectors are given by
\begin{eqnarray}
\label{eq:h3pts}
|R_\pm(\theta)\rangle=\frac{1}{2}\left[\begin{array}{c} e^{\pm i\theta}\\\pm\sqrt{2}\\e^{\mp i\theta}\end{array}\right], && 
|R_0(\theta)\rangle=\frac{1}{\sqrt{2(1+\sin^2\theta)}}\left[\begin{array}{c}-1\\\sqrt{2}i\sin\theta\\1\end{array}\right].
\end{eqnarray}

We leave it to the reader to check that (i) the right eigenvectors are not orthogonal; (ii) the left and right eigenvectors satisfy $\langle L_k|R_j\rangle=0$ when $k\neq j$; (iii) the left-right inner products $\langle L_k|R_k\rangle$ are not unity, and (iv) the resolution of identity is given by 
\begin{equation}
\label{eq:i3}
\frac{|R_+\rangle\langle L_+|}{\langle L_+|R_+\rangle}+\frac{|R_-\rangle\langle L_-|}{\langle L_-|R_-\rangle}+\frac{|R_0\rangle\langle L_0|}{\langle L_0|R_0\rangle}=\mathbbm{1}_3.
\end{equation}

When $\gamma/J=\cosh\beta\geq 1$, spectrum of $H_3$ is given by $\{\pm i\sinh\beta,0\}$ and the corresponding Dirac-normalized, right eigenvectors are given by 
\begin{eqnarray}
\label{eq:h3ptb}
|R_\pm(\beta)\rangle=\frac{1}{\sqrt{2(1+\cosh2\beta)}}\left[\begin{array}{c} -e^{\pm\beta}\\\sqrt{2}i\\e^{\mp\beta}\end{array}\right], && 
|R_0(\beta)\rangle=\frac{1}{\sqrt{2(1+\cosh^2\beta)}}\left[\begin{array}{c}-1\\\sqrt{2}i\cosh\beta\\1\end{array}\right].
\end{eqnarray}
These results for eigenvectors allow us to explicitly construct the intertwining operators, Eqs.(\ref{eq:eta10})-(\ref{eq:eta12}) when the spectrum is purely real or has a complex-conjugate pair. For both cases, we leave as an exercise to the reader to check that the three $\eta$ matrices obtained in this manner satisfy the intertwining relation, Eq.(\ref{eq:eta2}). 

We see that the conserved eigenvector projections generalize naturally from real to complex-conjugate eigenvalues but no further. This provides a fixed limit of existence for those quantities, and shows that their existence is not so ubiquitous as it may have looked on the first sight. In our formalism, the building blocks of a general $\eta$ are projections, and therefore all $n$ of them would be needed to result into an invertible operator, just as expected. However, our results hold more generally even for Hamiltonians having one or more dissipation channels. Such Hamiltonians may still admit some conserved operators, though not as many as a Hamiltonian with full rank.


\subsection{Spectrum with diabolic-point degeneracies}
\label{subsec:lcross}

When the spectrum of $H\neq H^\dagger$ becomes degenerate, the sufficiency part in the above proof ceases to hold, and there might exist more conserved quantities than those found explicitly so far. Their exact count depends on the nature of the degeneracy. For eigenvalue-only or diabolic-point degeneracy, we now show that it's possible to explicitly construct additional independent conserved quantities in a very simple fashion. 

Let the $k_D$-dimensional subspace associated with an eigenvalue $\epsilon_k$ be spanned by $k_D$ orthonormal, right-eigenvectors $\{|R_k,a\rangle|a=1,\cdots,k_D\}$, with the corresponding set of the left-eigenvectors denoted by $\{\langle L_k,a|\}$. It is easy to check that the operators 
\begin{equation}
\label{eq:etaab}
\eta_k^{ab}=|L_k,a\rangle\langle L_k,b|
\end{equation}
are also conserved. We note that the $\eta_k^{ab}$ defined in Eq.(\ref{eq:etaab}) are not Hermitian; however, they can be used to create symmetric and antisymmetric Hermitian combinations. Therefore, the total number of independent, conserved, Hermitian operators becomes $\sum_k  k_D^2$. As is expected, this expression turns out to be equal to $n$ when the spectrum is non-degenerate. Also, it is equal to $n^2$ when it is maximally degenerate, i.e. the Hamiltonian is an identity matrix and trivially conserves any operator whatsoever. We stress that this result is purely algebraic, and does not rely on the existence of special symmetries for the underlying model.


\subsection{Spectrum with exceptional-point degeneracies}
\label{subsec:ep}

The remaining scenario concerns a degeneracy where both eigenvalues and their underlying eigenvectors coalesce, i.e. EP degeneracies. As a result of the eigenvector coalescence, the spectral resolution approach used above ceases to work. We can however bring those ideas back in a slightly different form using the notion of generalized right eigenvectors. These are vectors $|v_{Rm}\rangle$ that satisfy the equation $(H-\lambda)^m |v_{Rm}\rangle = 0$ for $m\geq 2$. Note that at $m=1$, this definition gives the regular right eigenvector at the EP. The set of generalized eigenvectors always form a basis of the underlying Hilbert space. Moreover, the spectral resolution results can be extended to non-diagonalizable Hamiltonians at an EP by using additional terms of the form $|v_{Rm}\rangle\langle v_{Lm}|$ for each generalized eigenvector of the Hamiltonian. Therefore, for an $n$-dimensional Hamiltonian, an exceptional point of order $N\leq n$ gives rise to $N$ conserved operators, in addition to the $n-N$ operators that are generated by eigenvectors corresponding to the non-degenerate eigenvalues. 

As an example, let us consider $H_3$ at the EP. At this point, the sole right-eigenvector satisfies the equation $H_3|v_{R1}\rangle=0$ and is given by $|v_{R1}\rangle=(-1,\sqrt{2}i,1)^T/2$. We also note that 
\begin{equation}
\label{eq:epv1}
|v_{R1}\rangle=\lim_{\theta\rightarrow\pi/2}|R_{\pm,0}(\theta)\rangle=\lim_{\beta\rightarrow 0}|R_{\pm,0}(\beta)\rangle.
\end{equation}
The second (generalized) eigenvector $|v_{R2}\rangle$ is defined by the equation $H_3^2|v_{R2}\rangle=0$. A convenient method to define $|v_{R2}\rangle$ is by equation $H_3|v_{R2}\rangle=|v_{R1}\rangle$, and leads to $|v_{R2}\rangle=i(1,0,1)^T/2$. The third vector is defined by $H_3^3|v_{R3}\rangle=0$ and can be chosen as any vector that is linearly independent of $|v_{R1}\rangle$ and $|v_{R2}\rangle$. In each case, the left-eigenvectors are given by the transpose of the right ones. 

Thus, an $n$-dimensional Hamiltonian whose spectrum is either non-degenerate or has only EP degeneracies always admits $n$ conserved, intertwining operators, no matter how many and what order the EPs are. The fact that there are exactly $n$ linearly independent intertwining operators is quite surprising and shows that the number of conserved, independent operators is a stable quantity, governed by simple algebraic laws. This is because EP degeneracies are far more common than diabolic-point (level crossing) degeneracies. We also note that the conserved operators themselves remain smooth across an exceptional point. 


\section{Recursive procedure for generating conserved operators}
\label{sec:recursive}

The spectral-decomposition approach provides explicit, analytical formulae for constructing all conserved operators of a given Hamiltonian in terms of its eigenvectors. Unfortunately, that is not of much practical help because obtaining analytical expressions for eigenvectors is a highly nontrivial task even for small dimensions $n\sim 6$. However, we will show that there is a simple but powerful workaround to this issue: the conserved operators may be constructed iteratively by using the prescription
\begin{equation}
\eta_{k+1} = \eta_k H
\end{equation}
starting from an initial $\eta_1$ that is determined from the outset. Note that if $\eta_k=\eta_k^\dagger$ is Hermitian, the intertwining relation Eq.(\ref{eq:eta2}) implies that recursively obtained $\eta_{k+1}$ is also Hermitian. It also follows that the different $\eta$ operators do not commute with each other, and the commutator is proportional to the anti-Hermitian part of $H$. Finally, since $H$ obeys a characteristic polynomial equation of order $n$, it follows that $\eta_{N+1}=\eta_1 H^N$ can be written as a linear combination of lower-order operators $\eta_{k\leq N}$. Thus, for an $n$-dimensional system, this recursive procedure gives rise to $n$ non-commuting, linearly independent conserved observables~\cite{Bian2020}. 

For a broad class of transpose-symmetric Hamiltonians $\eta_1$ is found as follows.  For conserved quantities to exist,  the Hamiltonian must have purely real or complex-conjugate eigenvalues, i.e. an antilinear operator $\mathcal{A}$ commutes with it. Let us write $\mathcal{A}=\mathcal{PT}=\mathcal{L}*$  where $\mathcal{L}$ denotes {\it the entire linear part of} $\mathcal{A}$ and $*$ is purely the complex-conjugation operation. (In general, the operator $\mathcal{T}$ has a nontrivial unitary part; that has been absorbed into the operator $\mathcal{L}$.) It is easy to see that if $H=H^T$, then $\mathcal{L}$ is an intertwining operator. Once $\eta_1=\mathcal{L}$ is identified, the rest are determined by the recursive procedure. 

To illustrate the simplicity and power of this method, let us consider the Hamiltonian $H_3$. It follows that the three, linearly independent intertwining operators are given by 
\begin{align}
\eta_1&=\mathcal{P}_3,\\
\eta_2&=\mathcal{P}_3H_3,\\
\eta_3&=\mathcal{P}_3H^2_3.
\end{align}
The three-dimensional model and the subsequent analysis can be generalized $H_D=JS_x+i\gamma S_z$ where $S_x,S_z$ are $D$-dimensional representations of SU(2)~\cite{Graefe2008,Quiroz2019}. Although the spectral decomposition process becomes more and more cumbersome as $D$ increases, the recursive procedure maintains its simplicity. We emphasize that since Eq.(\ref{eq:eta2}) is a linear relation, the choice of intertwining operators is not unique. Any linear combination of $\{\eta_1,\ldots,\eta_D\}$s is also a conserved quantity. In general, the eigenvalues of $\eta_k$ are not positive definite. However, only in the $\mathcal{PT}$-symmetric region, one can create positive-definite $\eta$s that, in the literature, have been used to create a complex extension of quantum mechanics in the $\mathcal{PT}$-symmetric region~\cite{Bender2001,Mostafazadeh2002,Znojil2006}. 

We note that the transpose-symmetric-Hamiltonian constraint is satisfied by many models of balanced gain and loss systems. However, there are two major exceptions. The first is generalized Hatano-Nelson models that span classical, semiclassical, or quantum systems. The second is purely classical systems such as coupled mechanical or electrical oscillators, where the dynamics of (square-root of) the energy density is governed by a Hamiltonian with purely imaginary entries. Both cases are discussed in the following paragraphs.  


\subsection{Asymmetric Hatano Nelson model}
\label{sec:hnmodel}

Generalized Hatano-Nelson models are characterized by an asymmetric, nearest-neighbor tunneling profile~\cite{Hatano1996,Hatano1998}. They show non-Hermitian skin effect~\cite{LeeC2019}, generalized bulk-boundary correspondence~\cite{Helbig2020}, and lead to novel applications such as the topological funneling of light~\cite{Weidemann2020}. Here, we construct the first intertwining operator $\eta_1$, which then permits the recursive construction of remaining conserved observables. 

The prototypical Hatano-Nelson Hamiltonian with dimension $D$ is given by 
\begin{equation}
\label{eq:hn}
H_\mathrm{HN}=JS_x+i\gamma S_y\neq H^T_\mathrm{HN}.
\end{equation} 
Note that Eq.(\ref{eq:hn}) can either represent a small system with few sites, or alternatively it can represent the block-diagonal, momentum-space Hamiltonian for an infinite lattice with a unit cell of size $D$. The Hatano-Nelson model has asymmetric tunneling characterized by two energy scales $J\pm\gamma$ respectively. We remind the reader that for $D\geq 4$, the tunneling elements in $H_\mathrm{HN}$ depend on the site index. The Hamiltonian $H_\mathrm{HN}$, Eq.(\ref{eq:hn}), is equivalent to the traditional $\mathcal{PT}$-symmetric Hamiltonian $H_D$, 
\begin{equation}
\label{eq:hn2}
H_\mathrm{HN}=R_x(-\pi/2)H_DR_x(\pi/2),
\end{equation} 
due to the unitary transformation $R_x(-\pi/2)$ where $R_x(\theta)=\exp(-iS_x\theta)=R_x(-\theta)^\dagger$ is a rotation about the $x$-axis through angle $\theta$. Since the first intertwining operator for $H_D$ is known, $\eta_1=\mathcal{P}_D=\mathrm{antidiag}(1,\ldots,1)$, we can construct the one for the Hatano Nelson Hamiltonian as 
\begin{equation}
\label{eq:eta1hn}
\eta_1=R_x(-\pi/2)\mathcal{P}_DR_x(\pi/2),
\end{equation}
and the remaining $D-1$ operators then follow trivially. It is easy to check that when $D=2$, since the parity operator $\mathcal{P}_2=2\sigma_x$ commutes with the rotation operator, the so-called ``$\mathcal{PT}$-inner product''~\cite{Bender1999b,Bender2003} is a constant of motion for both $H_D$ and $H_\mathrm{HN}$. It is also important to note that for higher dimensions $D\geq 3$, the first intertwining operator $\eta_1$, Eq.(\ref{eq:eta1hn}), does not commute with the corresponding operator $\mathcal{P}_D$ for $H_D$. 


\subsection{Conserved quantities for $\mathcal{PT}$ electrical circuit models}
\label{sec:electric}

An inductor-capacitor ($LC$) circuit is the simplest electrical analog of a mass+spring pendulum where the energy oscillates between the kinetic form (inductor) and the potential form (capacitor). In contrast to the loss or gain in a mechanical system, it is easy to implement controlled loss and gain in electrical circuits by using resistors and operational amplifiers. Two such $LC$ oscillators, one with a resistor $R$ (loss) and second with negative resistor $-R$ (gain), form the loss and gain ``sites'' of a $\mathcal{PT}$-symmetric dimer. These ``sites'' can be coupled through mutual inductance~\cite{Schindler2011,Schindler2012,Wang2020}, a coupling capacitor~\cite{Chitsazi2017}, or a coupling inductor~\cite{LeonMontiel2018}.  

Let us consider two $LC$ circuits coupled through mutual inductance~\cite{Wang2020}. Kirchhoff laws govern the dynamics of voltages across capacitors $V_{1,2}(t)$ and currents across the inductors $I_{1,2}(t)$, and the resulting set of linear, coupled equations can be mapped onto Schrodinger-like equation for a state vector $|\psi(t)\rangle$ such that its norm is the instantaneous circuit energy, i.e. $\mathcal{E}(t)=\langle\psi(t)|\psi(t)\rangle$. The non-Hermitian Hamiltonian describing the dynamics of the real state vector $|\psi(t)\rangle$ is given by~\cite{Wang2020} 
\begin{equation}
\label{eq:ptelectric1}
\frac{H(\gamma)}{(\omega_0/2)}=i\left[\begin{array}{cccc}
-2\gamma & 0 & \gamma_0 & -\gamma_\mathrm{PT}\\
0 & 2\gamma & -\gamma_\mathrm{PT} & \gamma_0\\
-\gamma_0 & \gamma_\mathrm{PT} & 0 & 0\\
\gamma_\mathrm{PT}& -\gamma_0 & 0 & 0
\end{array}\right]=-\sigma_y\otimes(\gamma_0\mathbbm{1}_2-\gamma_\mathrm{PT}\sigma_x)-i\gamma(\mathbbm{1}_2+\sigma_z)\otimes\sigma_z.
\end{equation}

Here $\omega_0=1/\sqrt{LC}$ is the fundamental frequency of each oscillator, $\gamma=\omega_0(L/R)$ is the dimensionless measure of gain or loss, $\gamma_0=(1/\sqrt{1-\mu}+1/\sqrt{1+\mu})$ and $\gamma_\mathrm{PT}=(1/\sqrt{1-\mu}-1/\sqrt{1+\mu})$ denote the two values of $\gamma$ at which the Hamiltonian $H(\gamma)$ has an EP, and $0\leq \mu=M/L\leq 1$ denotes the dimensionless mutual inductance coupling between the gain and loss circuits. To obtain conserved quantities for this $\mathcal{PT}$ electric dimer, we need to obtain the first intertwining operator for $H(\gamma)\neq H^T(\gamma)$. Note that the Hamiltonian $H(\gamma)$ has the antilinear symmetry $\mathcal{A}=\mathcal{PT}$ with $\mathcal{P}=\mathbbm{1}_2\otimes\sigma_x$ and $\mathcal{T}=(\sigma_z\otimes\mathbbm{1}_2)*$. However, since $H(\gamma)$ is not transpose-symmetric, the linear part $\mathcal{L}$ of $\mathcal{A}=(\sigma_z\otimes\sigma_x)*$ is not an intertwining operator. 

Looking at the tensor product structure of Eq.(\ref{eq:ptelectric1}), it is clear that the source of the transpose asymmetry is the $\sigma_y$ term, and the unitary transformation $U=\exp(+i\pi\sigma_z/4)\otimes\mathbbm{1}_2$ will convert the Hamiltonian (\ref{eq:ptelectric1}) into a transpose-symmetric matrix $\mathbbm{H}=U HU^\dagger=\mathbbm{H}^T.$ Under the same transformation, the antilinear operator $\mathcal{A}$ maps into $\mathcal{A'}=U\mathcal{A}U^\dagger=\left[\exp(+i\pi\sigma_z/2)\sigma_z\otimes\sigma_x\right]*$. Here, we have used the complex conjugation property $U^*=U^\dagger$. Comparing the result with $\mathcal{A'}=\mathcal{L}'*$ gives $\mathbbm{1}_2\otimes\sigma_x$ as the first intertwining operator for  the matrix $\mathbbm{H}$, and thereby the first intertwining operator for $H(\gamma)$, Eq.(\ref{eq:ptelectric1}), is given by inverse $U$ transformation, i.e. 
\begin{equation}
\label{eq:ptelectric2}
\eta_1=\mathbbm{1}_2\otimes\sigma_x.
\end{equation}
The remaining operators are then trivially obtained through the recursive construction. We leave the analysis of two $LC$ circuits connected by a coupling inductor $L_c$~\cite{LeonMontiel2018} as an exercise to the reader. 


\section{Hamiltonians with $\mathcal{PT}$, anti-$\mathcal{PT}$, and other symmetries}
\label{sec:antipt}

\begin{figure}
\begin{subfigure}{.33\textwidth}
  \centering
  \begin{tikzpicture}[scale=0.65]
\begin{axis}[ axis lines=middle, xmin=-10, xmax=10, ymin=-10, ymax=10, xtick=\empty, ytick=\empty]
\addplot [only marks, red, mark options={scale=2}] table {
-10 0
-8  0
-5  0 
-4  0
7  0
3   0
};
\addplot [only marks, mark=o, blue, mark options={scale=2}] table {
-6  4
-6  -4
-3  2
-3  -2
2    1
2   -1
4   3
4   -3
};
\end{axis}
 \end{tikzpicture}
\caption{$\eta H=H^\dagger \eta$}
  \label{fig:sfig1}
\end{subfigure}
\begin{subfigure}{.33\textwidth}
  \centering
  \begin{tikzpicture}[scale=0.65]
\begin{axis}[
    axis lines=middle,
    xmin=-10, xmax=10,
    ymin=-10, ymax=10,
    xtick=\empty, ytick=\empty
]
\addplot [only marks, red, mark options={scale=2}] table {
-9 -9
-2  -2
-1  -1   
3  3
7  7
};
\addplot [only marks, mark=o, blue, mark options={scale=2}] table {
-4  -7
-7 -4
-2  -1
-1 -2
2   -3
-3  2
0   7
7  0
};
\addplot [domain=-10:10, samples=2, dashed] {1*x};
\end{axis}
\end{tikzpicture}
  \caption{$\eta H=e^{-i\pi/2}H^\dagger \eta$}
  \label{fig:sfig2}
\end{subfigure}
\begin{subfigure}{.33\textwidth}
  \centering
  \begin{tikzpicture}[scale=0.65]
\begin{axis}[
    axis lines=middle,
    xmin=-10, xmax=10,
    ymin=-10, ymax=10,
    xtick=\empty, ytick=\empty
]
\addplot [only marks, red, mark options={scale=2}] table {
0 0
};
\addplot [only marks, mark=o, blue, mark options={scale=2}] table {
-2  2
2 -2
-4  -1
4 1
1   -5
-1  5
3   -6
-3   6
};
\end{axis}
\end{tikzpicture}
  \caption{$\eta H=-H\eta$}
  \label{fig:sfig3}
\end{subfigure}
\caption{Schematic illustration of symmetries of eigenvalues for non-Hermitian Hamiltonians with different symmetries. (a) In a $\mathcal{PT}$-symmetric case, eigenvalues are reflection-symmetric about the horizontal axis, (b) For Hamiltonians with anyonic $\mathcal{PT}$-symmetry, they are reflection-symmetric about the dotted line with slope $-\tan(-\pi/4)=1$. (c) For Hamiltonians with a chiral symmetry, eigenvalues are reflection-symmetric through the origin, i.e. occur in pairs $(\epsilon_k,-\epsilon_k)$. Note that eigenvalues of anti-$\mathcal{PT}$-symmetric Hamiltonians (not shown) are  reflection-symmetric about the vertical axis.}
\end{figure}
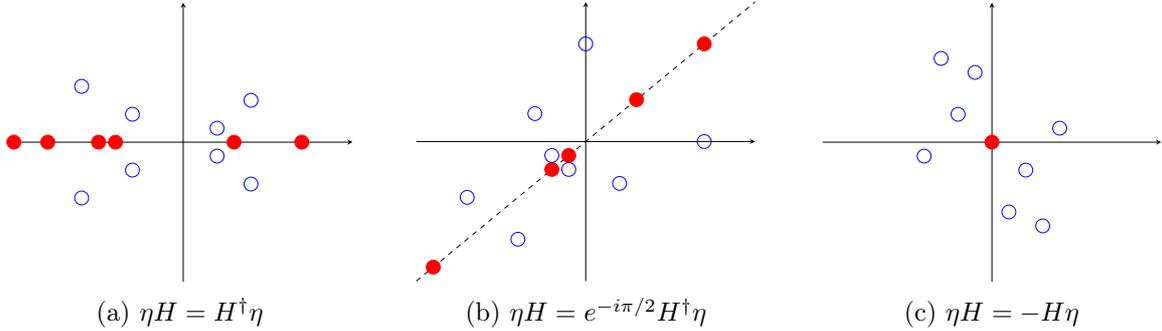  

The robust algebraic nature of the results about conserved observables for Hamiltonians with real or complex-conjugate spectra suggests several possibilities for further generalization~\cite{Simon2018,Alana2020}. The antilinear symmetry $\mathcal{A}=\mathcal{PT}$ that ensures real or complex-conjugate eigenvalues means the eigenvalues occur symmetrically about the $x$-axis in the complex plane (Fig.~\ref{fig:sfig1}). Here, we demonstrate that the results translate seamlessly from Schrodinger-like equations ($\mathcal{PT}$-symmetry) to diffusion equations (anti-$\mathcal{PT}$ symmetry) and beyond. 

To remind the reader about these ideas, we start with a diffusive system governed by equation $\partial_t |\psi(t)\rangle=\mathcal{D}|\psi(t)\rangle$. This can be written in the Schrodinger-like form with an $H_\mathcal{D}=i\mathcal{D}$. If the matrix $\mathcal{D}$ has an antilinear symmetry $\mathcal{A}$, it follows that the Hamiltonian $H_\mathcal{D}$ anticommutes with $\mathcal{A}$, i.e. $AH_\mathcal{D}=-H_\mathcal{D}A$. Such systems are called anti-$\mathcal{PT}$-symmetric systems, and their eigenvalues occur in pairs that are symmetric about the $y$-axis, i.e. they occur in pairs $(\epsilon_k,-\epsilon^*_k)$~\cite{Peng2016,Choi2018,LiY2019,Zhang2020,Wilkey2020}. Note that a Hamiltonian can be both $\mathcal{PT}$-symmetric and anti-$\mathcal{PT}$-symmetric with respect to different parity operations. The simplest example is a prototypical dimer with balanced gain and loss described by the $2\times2$ Hamiltonian $H_2=(J\sigma_x+i\gamma\sigma_z)/2$. We leave it to the reader to verify that $H_2$ is $\mathcal{PT}$ symmetric with respect to $\mathcal{PT}=\sigma_x*$, but {\it it is also anti-$\mathcal{PT}$-symmetric with respect to the operator $\mathcal{PT}=\sigma_z*$}. 

Following the procedure outlined in Sec.~\ref{sec:theory}, Hermitian conserved operators for a diffusive system can thus be generated recursively starting from the initial observable $\eta_1$ that satisfies a modified intertwining relation, 
\begin{align}
\label{eq:etaD}
\eta_1 H_\mathcal{D} &=-H_\mathcal{D}^\dagger\eta_1,\\
\eta_{k+1}&=i\eta_k H_\mathcal{D}=\eta_{k+1}^\dagger.
\end{align}

This recursive procedure generalizes to other systems where the equations of motion are linear, but do not fall neatly into either Schrodinger- or diffusion-equation categories. The conserved operators for such systems are defined by the intertwining relation 
\begin{equation}
\label{eq:etaphi}
\eta H=e^{i\varphi}H^\dagger\eta. 
\end{equation}
Physically, such Hamiltonians have eigenvalues that are symmetric about a line with slope $-\tan(\varphi/2)$ in the complex eigenvalue plane, i.e they occur in pairs $(\epsilon_k, e^{-i\varphi}\epsilon^*_k)$ (Fig.~\ref{fig:sfig2}). Equivalently, such a Hamiltonian satisfies the equation $\mathcal{A}H=e^{i\varphi}H\mathcal{A}$ for some antilinear operator $\mathcal{A}$. Note that $\varphi=0$ gives us the $\mathcal{PT}$-symmetric case whereas the anti-$\mathcal{PT}$-symmetric case corresponds to $\varphi=\pi$. Due to its similarity with quasiparticle-statistics definitions in two dimensions, Hamiltonians that require Eq.(\ref{eq:etaphi}) for their conserved observables are said to have ``anyonic $\mathcal{PT}$ symmetry''~\cite{Longhi2019,Arwas2021}. 

For an anyonic-$\mathcal{PT}$-symmetric Hamiltonian, we can reiterate the process of solving Eq.(\ref{eq:etaphi}) by spectrally decomposing all the operators in question. The recursive procedure for obtaining the constants of motion is also modified a bit,
\begin{equation}
\label{eq:ptanyon}
\eta_{k+1}=e^{i\varphi/2}\eta_kH=\eta_{k+1}^\dagger.
\end{equation} 
Finally, note that virtually the same behavior can be deducted by combination of equation-of-motion and other symmetries that the Hamiltonian may have. For example, if a Hamiltonian has a chiral symmetry, there exists an $\eta=\Pi$ that anticommutes with the Hamiltonian, i.e. $\eta H=-H\eta$. This, combined with the intertwining relation, implies that the spectrum of $H$ occurs in reflection-through-the-origin symmetric pairs $(\epsilon_k,-\epsilon_k)$ (Fig.~\ref{fig:sfig3}). Consequently, it gives rise to conserved pseudochirality in a simple, non-Hermitian model~\cite{Rivero2020}. 


\section{Discussion and Summary}
\label{sec:conc}

Now, we present the consequences of conserved quantities for lossy, non-Hermitian systems, briefly discuss generalizations of this approach to time-periodic or discrete-time systems, and then conclude with a summary. 


\subsection{Consequences for passive, $\mathcal{PT}$-symmetric systems}
\label{sec:lossypt}

$\mathcal{PT}$-symmetric Hamiltonians are realized in classical wave-systems where balanced gain and loss can be implemented by ignoring the thermal and quantum fluctuation effects~\cite{Purkayastha2020}. Conserved quantities in such systems lead to phase locking at the EP or in the $\mathcal{PT}$-broken region because the norm of an arbitrary state $\langle\psi(t)|\psi(t)\rangle$ grows with time whereas the expectation values $\langle\psi(t)|\eta_k|\psi(t)\rangle$ remain constant~\cite{Onanga2018}. But what about the quantum domain? 

The non-Hermitian Hamiltonians that arise for truly quantum systems through post-selection~\cite{Naghiloo2019,Quiroz2019} are always lossy. In such cases, the passive $\mathcal{PT}$-transition across the EP is marked by the emergence of a slowly decaying mode {\it whose decay-rate decreases when the local loss strength increases}~\cite{LeonMontiel2018,Joglekar2018}. Since the lossy Hamiltonian does not have complex-conjugate eigenvalues, in principle, the intertwining relation cannot be satisfied. However, the state norm $\langle\psi(t)|\psi(t)\rangle$ and the expectation values $\langle\psi(t)|\eta_k|\psi(t)\rangle$ decay at different rates in the passive $\mathcal{PT}$-symmetry broken region. Due to the exponential-in-time separation between these two quantities, the phase locking occurs for passive models~\cite{Bian2020}. These considerations show that all such effects continue be valid for {\it any non-Hermitian Hamiltonian that results from a $\mathcal{PT}$-symmetric Hamiltonian by a complex identity shift.}


\subsection{Conserved quantities in time-periodic models} 
\label{sec:dtqw}
Our analysis for conserved observables is applicable to a static Hamiltonian $H\neq H^\dagger$. In general, when the Hamiltonian $H(t)$ is time dependent, time-translational invariance is broken and so are the conservation laws. If the Hamiltonian is time periodic with period $T$, the time-translational symmetry is restored for discrete shifts $t\rightarrow t+pT$ where $p$ is an integer. The long-term dynamics of the system are then governed by the time evolution operator for one period, 
\begin{equation}
\label{eq:gfloquet1}
G_F(T)=\mathbbm{T} e^{-i\int_0^T dt' H(t')}\equiv e^{-iT H_F},
\end{equation}
where $\mathbbm{T}$ denotes the time-ordered product that accounts for the non-commuting nature of $H(t)$ at different instances of time. Note that Eq.(\ref{eq:gfloquet1}) also defines the (static) Floquet Hamiltonian $H_F$~\cite{Joglekar2014,Lee2015}. $G_F(T)$ determines the stroboscopic dynamics, namely dynamics only at integer times $pT$, while the dynamics during the intermediate times $pT\leq t\leq (p+1)T$, called the micromotion, is determined by the family of operators
\begin{equation}
K(t)=\mathbbm{T} e^{-i\int_0^t dt' H(t')}. 
\end{equation}

It is easy to show that if the Hamiltonian $H(t)$ is Hermitian, then so is the Floquet Hamiltonian $H_F$. Similarly, if the Hamiltonian $H(t)$ is $\mathcal{PT}$-symmetric, the resultant Floquet Hamiltonian $H_F$ also has an antilinear symmetry, albeit with possibly different $\mathcal{P}$ and $\mathcal{T}$ operators~\cite{Harter2020}. Under the stroboscopic evolution, an observable $\eta$ is conserved provided it satisfies
\begin{equation}
G^\dagger_F\eta G_F=\eta. 
\label{eq:dtqw1}
\end{equation}
Equation (\ref{eq:dtqw1}) defines the ``intertwining relation'' for systems that are described by a time-evolution operator $G$ instead of a Hamiltonian. In particular, it defines stroboscopically conserved quantities for non-Hermitian, discrete time quantum walks. The recursive procedure for generating other intertwining operators follows through with 
\begin{equation}
\label{eq:dtqw2}
\eta_{k+1}=\eta_k G.
\end{equation}
Thus, a complete characterization of conserved quantities for non-Hermitian, time-periodic Hamiltonians as well as non-unitary discrete time quantum walks can be carried out with the explicit, analytical, recursive procedure we have reviewed in this article. 


\subsection{Summary}
In this article, we have reviewed conserved quantities that arise in the dynamics of systems that are governed by non-Hermitian Hamiltonians with different antilinear symmetries. For isolated quantum systems governed by Hermitian Hamiltonians, the conserved observables are linearly independent operators that commute with the Hamiltonian. In contrast, for a quantum channel,  generically, there are no conserved quantities except the trace of the density matrix. Classical or quantum systems governed by effective, non-Hermitian Hamiltonians lie between these two limits. Non-Hermitian Hamiltonians with different antilinear symmetries ($\mathcal{PT}$, anti-$\mathcal{PT}$, anyonic-$\mathcal{PT}$) give rise to different intertwining relations that characterize conserved observables. 

The fact that eigenvector-projections of Hermitian operators remain conserved during time evolution is quite obvious, although often somewhat neglected when discussing conservation laws in quantum mechanics. With that insight, we have reviewed the spectral decomposition method for constructing conserved quantities for the non-Hermitian cases, and then outlined a simple recursive, analytical solution for a complete set of linearly independent intertwining operators. Note that each conserved operator $\eta_k$ generates a one-parameter unitary group $U_k(\theta)=\exp(i\theta\eta_k)$ by exponentiation. However, the unitary transformation does not leave the Hamiltonian invariant, i.e. $U_k(\theta)H\neq HU_k(\theta)$, and therefore it does not have a direct physical meaning.

In this article, we have further quantified the number of conserved observables for non-Hermitian Hamiltonians with either diabolic degeneracies (level crossings) or EP degeneracies, the latter being most commonplace for non-Hermitian matrices. While in the first case we discovered new conserved observables, we showed that that crossing through an EP degeneracy does not yield any new ones. Thus, in the complex eigenvalue plane, the number of conserved observables remains equal to the maximum rank of the Hamiltonian matrix. 


\section*{References}

\bibliographystyle{iopart-num}
\bibliography{ptJPCS.bib} 

\end{document}